\documentclass{ws-procs975x65}

\newcommand{\dd}[0]{\textrm{d}}
\newcommand{\defn}[0]{\equiv}

\newcommand{\qsubrm}[2]{{#1}_{\scriptscriptstyle{\textrm{#2}}}}

\newcommand{\qsuprm}[2]{{#1}^{\scriptscriptstyle\textrm{#2}}}

\newcommand{\pis}[0]{ {\Pi}^{\scriptscriptstyle\rm{S}}}

\newcommand{\rbm}[1]{{\bf{#1}}}
\newcommand{\ci}[0]{\textrm{i}}

\renewcommand{\figurename}{Figure}

\def\be{\begin{equation}}
\def\ee{\end{equation}}
\def\bea{\begin{eqnarray}}
\def\eea{\end{eqnarray}}
\def\bse{\begin{subequations}}
\def\ese{\end{subequations}}

\let\oldsqrt\sqrt
\def\sqrt{\mathpalette\DHLhksqrt}
\def\DHLhksqrt#1#2{%
\setbox0=\hbox{$#1\oldsqrt{#2\,}$}\dimen0=\ht0
\advance\dimen0-0.2\ht0
\setbox2=\hbox{\vrule height\ht0 depth -\dimen0}%
{\box0\lower0.4pt\box2}}

\def\beq{\begin{equation}}
\def\eeq{\end{equation}}

\begin{document}
\title{The Equation of State Approach to Cosmological Perturbations in $f(\mathcal{R})$ Gravity}

\author{Boris Bolliet}
\address{Laboratoire de Physique Subatomique et de Cosmologie, Universit\'e Grenoble-Alpes, CNRS/IN2P3\\
53, avenue des Martyrs, 38026 Grenoble cedex, France}
\author{Richard A. Battye}
\address{Jodrell Bank Centre for Astrophysics, School of Physics and Astronomy, The University of Manchester, Manchester M13 9PL, U.K.}
\author{Jonathan A. Pearson}
\address{School of Physics and Astronomy, University of Nottingham, Nottingham NG7 2RD, U.K.}
\begin{abstract}
The discovery of apparent cosmological acceleration has spawned a huge number of dark energy and modified gravity theories. The $f\left(\mathcal{R}\right)$ models of gravity are obtained when one replaces the Ricci scalar in the Einstein-Hilbert action by an arbitrary function $f\left(\mathcal{R}\right)$, 
\be
\label{eq:FR-action-1}
S = \frac{1}{2\kappa}\int \dd^4x\, \sqrt{-g} \,  \left\{  {\mathcal{R} + f\left(\mathcal{R}\right)} \right\}  + \qsubrm{S}{m},
\ee
where $\kappa\equiv 8\pi {G}$ is the rescaled  Newton's constant, $\mathcal{R}$ is the Ricci scalar and $\qsubrm{S}{m}$ is the action describing the standard matter fields. 
In this work we provide expressions for the equations of state (EoS) for perturbations which completely characterize the linearized perturbations in $f\left(\mathcal{R}\right)$ gravity, including the scalar, vector, and tensor modes. The EoS formalism \cite{Battye:2013aaa} is a powerful and elegant parametrization where the modification to General Relativity are treated as a dark-energy-fluid. The perturbed dark-energy-fluid variables such as the anisotropic stress or the entropy perturbation are explicitly given in terms of parameters of the model of interest. 
\end{abstract}
\bodymatter
\section{Field Equations}
Varying the action (\ref{eq:FR-action-1}) with respect to the metric yields the field equations,
\bea
\label{eq:U-gen-fr-full}
G_{\mu\nu} = \kappa \left( T_{\mu\nu} + D_{\mu\nu}\right),
\eea
where $G_{\mu\nu}$ is the Einstein tensor and $T_{\mu\nu}$ is the stress-energy tensor of the  standard matter fields. All contributions due to $f\left(\mathcal{R}\right)$  are packaged into the extra-term $D_{\mu\nu}$, which we call the stress-energy tensor of the dark sector, explicitly formulated as
\bea
\label{eq:U-gen-fr}
\kappa D_{\mu\nu} \defn  \tfrac{1}{2}fg_{\mu\nu}- \left( R_{\mu\nu}+g_{\mu\nu}\square - \nabla_{\mu}\nabla_{\nu}\right)f_{\mathcal{R}},
\eea
where $R_{\mu\nu}$ is the Ricci tensor, and $f_\mathcal{R}\equiv\frac{df}{d\mathcal{R}}$. A direct calculation shows that $D_{\mu\nu}$ is covariantly conserved, that is, $\nabla^{\mu}D_{\mu\nu}=0$. The background geometry is assumed to be $ds^2=-dt^2+a\left(t\right)^2\delta_{ij}dx^{i} dx^{j}$, where $a\left(t\right)$ is the scale factor. Instead of the first and second order time derivative of the Hubble parameter, $H$, we use the dimensionless parameters\footnote{With these notation the Ricci scalar reads $\mathcal{R}=12H^2(1-\tfrac{1}{2}\qsubrm{\epsilon}{H})$.}
\be
\qsubrm{\epsilon}{H}\equiv- \frac{H^\prime}{H},\,\,\,\,\,\,\,\,\,\,\,\,\,\,\qsubrm{\bar{\epsilon}}{H}\equiv-\frac{\mathcal{R}^\prime}{6H^2},
\ee
where the prime denotes derivative with respect to $d/d\ln a$. The dark sector can be viewed as a fluid, with energy density $\qsubrm{\rho}{de} \equiv \tfrac{1}{a^2} D_{00}$ and pressure $\qsubrm{P}{de} \equiv \tfrac{1}{3a^2} \delta^{ij}D_{ij}$. The field equations (\ref{eq:U-gen-fr-full}) read
\be
\qsubrm{\Omega}{m}+\qsubrm{\Omega}{de}=1,\,\,\,\,\,\,\,\,\,\,\,\,\,\,\qsubrm{w}{m}\qsubrm{\Omega}{m}+\qsubrm{w}{de}\qsubrm{\Omega}{de}=\tfrac{2}{3}\qsubrm{\epsilon}{H} -1,\label{eq:fe}
\ee
where $\qsubrm{\Omega}{i}=\frac{\kappa\qsubrm{\rho}{i}}{3H^{2}}$ is the density parameter and $\qsubrm{w}{i}\equiv \qsubrm{P}{i}/\qsubrm{\rho}{i}$. From (\ref{eq:U-gen-fr}). For the $f(\mathcal{R})$ fluid, they are explicitly given by
\bse
\begin{eqnarray}
\qsubrm{\Omega}{de} & = & -\frac{f}{6H^2}+(1-\qsubrm{\epsilon}{H})f_\mathcal{R}-f_\mathcal{R}^\prime,\label{eq:exrho}\\
\qsubrm{w}{de}+1  & = & -\frac{1}{3\qsubrm{\Omega}{de}}\left(2\qsubrm{\epsilon}{H}f_\mathcal{R}+(1+\qsubrm{\epsilon}{H})f_\mathcal{R}^\prime-f_\mathcal{R}^{\prime\prime}\right).\label{eq:exw}
\end{eqnarray}
\ese
\section{Dynamics of Linear Perturbations}
The dynamics of linear perturbations is presented in Fourier space. Instead of the coordinate wavenumber, $k$, a reduced dimensionless wavenumber is introduced, 
\bea
\mathrm{K}\equiv\frac{k}{aH},
\eea
so that $K$ can easily recognize the sub-(super)-horizon regimes. In the synchronous gauge, the non-zero metric perturbations are $\delta g_{ij}=a^2 h_{ij}$. In an orthonormal basis $\{\hat k, \hat l, \hat m\}$ in $k$-space, the spatial matrix $h_{ij}$ is further decomposed as 
$h_{ij}=\frac{1}{3}h\delta_{ij}+h_{\parallel}\sigma_{ij}+\qsuprm{h}{V}\cdot v_{ij}+\qsuprm{h}{T}\cdot e_{ij}$, where the notations $\qsuprm{h}{V}$ and $\qsuprm{h}{T}$ contain the two vector and the two tensor polarization states respectively, and the dot product has to be understood as a sum over the polarization states.  Instead of $h$, we use the combination $6\eta\equiv h_{\parallel}-h$. The basis matrices are $\sigma_{ij} =\hat{k}_{i}\hat{k}_{j}-\frac{1}{3}\delta_{ij}$ for the longitudinal traceless mode, $\qsuprm{v}{(1)}_{ij}=2\hat{k}_{(i}\hat{l}_{j)}$ and $\qsuprm{v}{(2)}_{ij}=2\hat{k}_{(i}\hat{m}_{j)}$ for the vector modes and $e_{ij}^{\times}=2\hat{l}_{[i}\hat{m}_{j]}$, $e_{ij}^{+}=\hat{l}_{i}\hat{l}_{j}-\hat{m}_{i}\hat{m}_{j}$ for the tensor modes. In the conformal Newtonian gauge,  $\delta{g_{00}}=-2a^{2}\psi$ and t $\delta{g_{ij}}=-2a^{2}\phi\delta_{ij}$ (the tensor and vector modes remain the same in both gauges). An additional scalar degree of freedom arises at the perturbative level from the non-vanishing $f_\mathcal{R}^\prime$, given by\footnote{With the gauge invariant notation the perturbed Ricci scalar reads $\delta\mathcal{R}=-6H^2(W+4X-\tfrac{1}{3}K^2(Y-2Z)-\qsubrm{\bar{\epsilon}}{H}T)$. The fact that $T$ appears explicitly in the expression of $\delta\mathcal{R}$ indicates that this is not a gauge invariant quantity.\label{fn:deltaR}}
\be
\chi\equiv-\frac{f_\mathcal{R}^\prime}{\qsubrm{\bar{\epsilon}}{H}}\frac{\delta\mathcal{R}}{6H^2}.\label{eq:chi}
\ee
This feature is a manifestation of the well-known connection between $f(\mathcal R)$ theories and non-minimally coupled scalar-tensor theories. 
The expression of a generic perturbed stress-energy tensor, ${U^{\mu}}_{\nu}$, is
\begin{equation}
{\delta U^{\mu}}_{\nu}=\left(\rho\delta+\delta P\right)u^{\mu}u_{\nu}+\left(\rho+P\right)\left(u_{\nu}\delta u^{\mu}+u^{\mu}\delta u_{\nu}\right)+\delta P{\delta^{\mu}}_{\nu}+P{{\Pi}^{\mu}}_{\nu},
\label{eq:pset}
\end{equation}
where the density contrast is $\delta \equiv \delta \rho/\rho$, the Hubble flow is parametrized by $u_\nu=(-1,\vec{0})$ in coordinate time, and $\delta u_\nu=(0,\delta u_i)$ is the perturbed velocity field whose scalar mode is $\theta  \equiv  \frac{\ci k^{j}\delta u_{j}}{k^{2}}$.

Our results are are obtained in both the synchronous and conformal Newtonian gauges thanks  to a new set of variables presented bellow (quantities denoted with the subscript `S' (`C') are evaluated in the synchronous (conformal Newtonian) gauges respectively).
\begin{eqnarray}
\begin{array}{ccc}
\quad \rbm{Symbol\quad} &\quad \rbm{Synchronous\,\,gauge\quad} & \quad\rbm{Conformal\,\,Newtonian\,\,gauge\quad}\\
T & \frac{{h}_{\parallel}^\prime}{2\rm{K}^{2}} & 0\\
Y & T^\prime+\qsubrm{\epsilon}{H}T & \psi\\
Z & \eta-T & \phi\\
X & Z^\prime+Y & Z^\prime+Y\\
W &X^\prime-\qsubrm{\epsilon}{H}(X+Y) & X^\prime-\qsubrm{\epsilon}{H}(X+Y)\\
\hat{\Theta} & \Theta_{s}+3\left(1+w\right)T & \Theta_{c}\\
\hat{\delta P} & \delta P_{s}+P_{s}^\prime T & \delta P_{c}\\
\hat{\chi} & \chi_{s}+f_\mathcal{R}^\prime T &  \mathcal{\chi}_{c}
\end{array}
\end{eqnarray}
Instead of $\delta$ and $\theta$, we make an extensive use of the  dimensionless variables
\be
\Delta \equiv \delta+3\left(1+w\right)H\theta, \,\,\,\,\,\,\,\,\,\,\,\,\,\,\Theta \equiv 3\left(1+w\right)H\theta.
\ee
The perturbed pressure is packaged into the gauge invariant entropy perturbation,
\begin{equation}
w\Gamma  =  \frac{\hat{\delta P}}{\rho}-\frac{dP}{d\rho}\left(\Delta-\hat{\Theta}\right)\label{eq:gammagi}.
\end{equation}
The anisotropic stress is the spatial traceless part of the stress-energy tensor.  In the same way as the metric perturbation, it decomposes into one scalar, $\pis$, two vector, $\qsuprm{\Pi}{V}$, and two tensor modes, $\qsuprm{\Pi}{T}$.  
The generic perturbed fluid equation, $\delta(\nabla^\mu U_{\mu\nu})=0$, are 
\bse
\label{eq:pfe}
\begin{eqnarray}
{\Delta^\prime}-3w\Delta-2w\qsuprm{\Pi}{S}+\qsubrm{g}{K}\qsubrm{\epsilon}{H}\hat{\Theta} & = & 3\left(1+w\right)X,\label{eq:deltadot}\\
{\hat{\Theta}}^{\prime}+\left(\qsubrm{\epsilon}{H}-\frac{{w^{\prime}}}{1+w}\right)\hat{\Theta}-\left(3w-\frac{w^{\prime}}{1+w}\right)\Delta-2w\qsuprm{\Pi}{S}-3w\Gamma & = & 3\left(1+w\right)Y.\label{eq:Thetadot}
\end{eqnarray}
\ese
The prime denote derivative with respect to $d/d\ln a$ and $\qsubrm{g}{K}\equiv1+\frac{\rm{K}^2}{3\qsubrm{\epsilon}{H}}$. The field equations (\ref{eq:U-gen-fr-full}) expanded to linear order in perturbations are
\bse
\label{eq:dgdu}
\begin{eqnarray}
-\tfrac{2}{3}\mathrm{K}^{2}Z & = & \qsubrm{\Omega}{m}\qsubrm{\Delta}{m}+\qsubrm{\Omega}{de}\qsubrm{\Delta}{de},\label{eq:deltam}\\
2X & = & \qsubrm{\Omega}{m}\qsubrm{\hat{\Theta}}{m} + \qsubrm{\Omega}{de}\qsubrm{\hat{\Theta}}{de},\label{eq:Thetam}\\
\tfrac{2}{3}W+2X-\tfrac{2}{9}\mathrm{K}^{2}\left(Y-Z\right) & = &\qsubrm{\Omega}{m}(\qsubrm{\hat{\delta P}}{m}/\qsubrm{\rho}{m}) +  \qsubrm{\Omega}{de}(\qsubrm{\hat{\delta P}}{de}/\qsubrm{\rho}{de}),\label{eq:deltaP_m}\\
\tfrac{1}{3}\mathrm{K}^{2}\left(Y-Z\right) & = &\qsubrm{\Omega}{m}\qsubrm{w}{m}\qsubrm{\Pi}{m}^\mathrm{S} +  \qsubrm{\Omega}{de} \qsubrm{w}{de}\qsubrm{\Pi}{de}^\mathrm{S},\label{eq:Pi_Sm}\\
\tfrac{1}{6}{\qsuprm{h}{V}}^{\prime\prime}+(\tfrac{1}{2}-\tfrac{1}{6}\qsubrm{\epsilon}{H}){\qsuprm{h}{V}}^\prime & = &\qsubrm{\Omega}{m}\qsubrm{w}{m}\qsubrm{\Pi}{m}^\mathrm{V}+  \qsubrm{\Omega}{de} \qsubrm{w}{de}\qsubrm{\Pi}{de}^\mathrm{V},\label{eq:Pi_V_m}\\
\tfrac{1}{6}{\qsuprm{h}{T}}^{\prime\prime}+(\tfrac{1}{2}-\tfrac{1}{6}\qsubrm{\epsilon}{H}){\qsuprm{h}{T}}^\prime+\tfrac{1}{3}\mathrm{K}^{2}\qsuprm{h}{T} & = & \qsubrm{\Omega}{m}\qsubrm{w}{m}\qsubrm{\Pi}{m}^\mathrm{T} +  \qsubrm{\Omega}{de} \qsubrm{w}{de}\qsubrm{\Pi}{de}^\mathrm{T}.\label{eq:Pi_T_m}
\end{eqnarray}
\ese
\section{Equation of State for Perturbations}
In the $f({\mathcal{R}})$ scenario, the expansion to first order in perturbations of the dark sector stress-energy tensor is 
\begin{eqnarray}
\kappa\delta D_{\mu\nu} & = & -f_{\mathcal{R}}\delta R_{\mu\nu}+\tfrac{1}{2}f\delta g_{\mu\nu}+\tfrac{1}{2}g_{\mu\nu}f_{\mathcal{R}}\delta\mathcal{R}-f_{\mathcal{RR}}R_{\mu\nu}\delta\mathcal{R}\nonumber \\
 &  & \quad +\delta\left(\nabla_{\mu}\nabla_{\nu}f_{\mathcal{R}}\right)-\left(\square f_{\mathcal{R}}\right)\delta g_{\mu\nu}-g_{\mu\nu}\delta\left(\square f_{\mathcal{R}}\right).\label{eq:pee}
\end{eqnarray}
This allows us to isolate the perturbed fluid variables for the $f(\mathcal{R})$ dark sector theory. The tensor and vector projections of (\ref{eq:pee}) readily constitute the EoS for $\qsubrm{\Pi}{de}^\mathrm{V}$ and $\qsubrm{\Pi}{de}^\mathrm{T}$,
\bse
\label{eq:dudefvt}
\begin{eqnarray}
\qsubrm{\Omega}{de} \qsubrm{w}{de}\qsubrm{\Pi}{de}^\mathrm{V} & = & -\tfrac{1}{6}f_{\mathcal{R}} {\qsuprm{h}{V}}^{\prime\prime}-\tfrac{1}{6}\left\{(3-\qsubrm{\epsilon}{H})f_{\mathcal{R}}+f_{\mathcal{R}}^\prime\right\}{\qsuprm{h}{V}}^\prime ,\label{eq:Pi_V}\\
 \qsubrm{\Omega}{de} \qsubrm{w}{de}\qsubrm{\Pi}{de}^\mathrm{T} & = & -\tfrac{1}{6}f_{\mathcal{R}} {\qsuprm{h}{T}}^{\prime\prime}-\tfrac{1}{6}\left\{(3-\qsubrm{\epsilon}{H})f_{\mathcal{R}}+f_{\mathcal{R}}^\prime\right\}{\qsuprm{h}{T}}^\prime-\tfrac{1}{6}f_{\mathcal{R}}\mathrm{K}^{2}\qsuprm{h}{T}. \label{eq:Pi_T}
\end{eqnarray}
\ese
The time-time projection provides a useful formulae that enables to write $\hat{\chi}$ in terms of $\qsubrm{\Delta}{de},X$ and $Z$,
\be
\qsubrm{\Omega}{de}\qsubrm{\Delta}{de}  =  -\qsubrm{g}{K}\qsubrm{\epsilon}{H}\hat{\chi}+f_{\mathcal{R}}^\prime(X+\tfrac{1}{3}\mathrm{K}^{2}Z).\label{eq:delta}
\ee
The longitudinal spatial traceless projection is
\be
\qsubrm{\Omega}{de} \qsubrm{w}{de}\qsubrm{\Pi}{de}^\mathrm{S}  =  -\tfrac{1}{3}\mathrm{K}^{2}\hat{\chi}-\tfrac{1}{3}f_{\mathcal{R}}\mathrm{K}^{2}\left(Y-Z\right).\label{eq:Pi_S}
\ee
Using (\ref{eq:Pi_S},\ref{eq:delta}) and the perturbed field equations (\ref{eq:dgdu}), it is possible to write $\qsubrm{\Gamma}{de}$ and $\qsubrm{\Pi}{de}^\mathrm{S}$ in terms of the other perturbed fluid variables:
\bse
\label{eq:EoS}
\begin{eqnarray}
\qsubrm{w}{de}\qsubrm{\Gamma}{de} & = & (\qsubrm{\zeta}{de}-\tfrac{2}{3}\tfrac{\qsubrm{\bar{\epsilon}}{H}}{\qsubrm{g}{K}\qsubrm{\epsilon}{H}}\tfrac{1}{f_\mathcal{R}^\prime})\qsubrm{\Delta}{de}-\qsubrm{\zeta}{de}\qsubrm{\hat{\Theta}}{de}+\tfrac{\qsubrm{\Omega}{m}}{\qsubrm{\Omega}{de}}\qsubrm{\zeta}{m}(\qsubrm{\Delta}{m}-\qsubrm{\hat{\Theta}}{m})\label{eq:eosGamma}\\
\qsubrm{w}{de}\qsubrm{\Pi}{de}^\mathrm{S} & = & \tfrac{\qsubrm{g}{K}-1}{\qsubrm{g}{K}}\tfrac{1}{1+f_\mathcal{R}}\left\{(1+\tfrac{1}{2}f_\mathcal{R}^\prime)\qsubrm{\Delta}{de}-\tfrac{1}{2}f_\mathcal{R}^\prime\qsubrm{\hat{\Theta}}{de}+\tfrac{1}{2}\tfrac{\qsubrm{\Omega}{m}}{\qsubrm{\Omega}{de}}f_\mathcal{R}^\prime(\qsubrm{\Delta}{m}-\qsubrm{\hat{\Theta}}{m})\right\}\label{eq:eosPi_S}
\end{eqnarray}
\ese
where $\qsubrm{\zeta}{i} \equiv\frac{\qsubrm{g}{K}\qsubrm{\epsilon}{H}-\qsubrm{\bar{\epsilon}}{H}}{3\qsubrm{g}{K}\qsubrm{\epsilon}{H}}-\frac{\qsubrm{d{P}}{i}}{\qsubrm{d{\rho}}{i}}$. These expressions constitute the EoS for perturbations in $f(R)$ gravity.
From here, the whole dynamics of the (scalar) perturbations is provided by the four perturbed fluid equations (\ref{eq:pfe}), plus one evolution equation for the metric perturbations\footnote{$Z^\prime=X-Y$, where $X$ and $Y$ are written in terms of the perturbed fluid variables.}\footnote{In fact, the system of five differential equations is overdetermined: when $K=0$, the field equation (\ref{eq:deltam}) can be used to express $\qsubrm{\Delta}{de}$ in terms of $\qsubrm{\Delta}{m}$ ; and when $K\neq0$, the same equation gives $Z$ in terms of the  $\qsubrm{\Delta}{i}$'s}. This approach is as powerful as elegant: it provides an efficient way to solve the linear perturbation in $f(R)$ gravity, and the phenomenology becomes transparent through the interpretation of the fluid variables. 
Here we have illustrated the EoS approach with f(R) gravity, but it should apply to any modified gravity theory.
The EoS approach can be an alternative (or complementary) to the ``parametrized post Friedmann framework" developped by Ferreira-Baker-Skordis. It enables to simply characterize intricate modified gravity theories through their equations of state for perturbations.
In the quest for understanding the nature of dark energy, this approach seems to us particularly attractive as it links formal modification to General Relativity to phenomenology in a straightforward way\cite{Battye:2016aaa}.

\end{document}